\documentclass[iop]{emulateapj}
\usepackage{amsmath}
\usepackage{color}

\shorttitle{}
\shortauthors{}
\begin{document}

\title{Eclipsing Stellar Binaries in the Galactic Center}
\author{Gongjie Li \altaffilmark{1, 2}, Idan Ginsburg \altaffilmark{1}, Smadar Naoz \altaffilmark{3, 4}, Abraham Loeb \altaffilmark{1}}
\affil{$^1$ Harvard-Smithsonian Center for Astrophysics, The Institute for Theory and
Computation, \\60 Garden Street, Cambridge, MA 02138, USA} 
\affil{$^2$ Center for Relativistic Astrophysics, School of Physics, Georgia Institute of Technology, Atlanta, GA 30332, USA }
\affil{$^3$ Department of Physics and Astronomy, University of California, Los Angeles, CA 90095, USA}
\affil{$^4$ Mani L. Bhaumik Institute for Theoretical Physics, Department of Physics and Astronomy, UCLA, Los Angeles, CA 90095, USA}
\email{gli@cfa.harvard.edu}

\begin{abstract}
Compact stellar binaries are expected to survive in the dense environment of the Galactic Center. The stable binaries may undergo Kozai-Lidov oscillations due to perturbations from the central supermassive black hole (Sgr A*), yet the General Relativistic precession can suppress the Kozai-Lidov oscillations and keep the stellar binaries from merging. However, it is challenging to resolve the binary sources and distinguish them from single stars. The close separations of the stable binaries allow higher eclipse probabilities. Here, we consider the massive star SO-2 as an example and calculate the probability of detecting eclipses, assuming it is a binary. We find that the eclipse probability is $\sim 30-50\%$, reaching higher values when the stellar binary is more eccentric or highly inclined relative to its orbit around Sgr A*.
\bigskip
\end{abstract}

%\keywords{Earth--chaos--instabilities, Exoplanets--dynamics}

\section{Introduction}
Observations of the inner region of our Milky Way provide valuable information on the stellar population and the supermassive black hole (SMBH), Sgr A*, in the Galactic Center, and allow a unique opportunity to study the interactions of the SMBH and its surrounding stars. There are three detected stellar binaries within $\sim 0.2$ pc of the Galactic Center: IRS16W \citep{Ott99, Martins06}, IRS 16NE and E60 \citep{Pfuhl13}. IRS16W and E60 are eclipsing binaries and IRS 16NE is a long period spectroscopic binary. Based on these three detections, it has been estimated that the binary fraction in the Galactic Center is similar to that of a young stellar cluster \citep{Pfuhl13}. 

The binary population play important roles in the dynamical processes at the Galactic Center. For instance, the binaries contribute to the relaxation of the Galactic Center \citep{Alexander09},  and they explain the origin of the young stellar population in the galactic center, such as the S-stars \citep{Antonini13}, the hypervelocity stars \citep[e.g.,][]{Hills88, Yu03, Ginsburg07, Perets09} and the G1 and G2 clouds \citep{Witzel14, Stephan16, Witzel17}. Moreover, the existence of a binary population may yield an underestimation of the GC stellar disk membership \citep{Naozinprep}.

Long term stability requires that binaries in the GC have a tight configuration in reference to their orbit around the SMBH. In particular, Kozai-Lidov oscillations \citep{Kozai62, Lidov62} due to perturbations from Sgr A* may excite the eccentricity of a stellar binary, reduce its pericenter distances, and lead to mergers of the binary components in the Galactic Center \citep{Prodan15, Naoz16, Stephan16}. \citet{Stephan16} found that within $0.1$ pc of the SMBH, $\sim 13\%$ of the initial binary population will merge in a few million years and $\sim 17\%$ will become unbound due to interactions with single stars. Moreover, $70\%$ of the initial population will remain bound to a binary companion after a few million years. 

%For the S-stars, \citet{Ginsburginprep} studied the parameter space at which binaries in the S-star cluster are expected to survive under the extreme gravitational perturbations from the SMBH (i.e., the  eccentric Kozai-Lidov mechanism \citealt{Naoz16}). General relativistic precession can suppress the eccentricity excitation caused by Kozai-Lidov oscillations, and stop the stellar binaries from merging \citep[e.g.,][]{Naoz13, Li15}. In particular,  \citet{Ginsburginprep} found that a stellar binary with semi-major axis $a_1<0.1$ AU is stable at the location of S2.

It is not clear whether the observed stellar systems in the Galactic Center are single stars or binaries, due to the high resolution required to distinguish between these possibilities. The existence of S-stars poses challenges regarding their formation, as the strong tidal field from Sgr A* should inhibit the collapse and fragmentation of the parent molecular cloud \citep{Morris93}. Thus, migration of the stars after formation may be involved in such systems \citep{Merritt09, Perets10}. The challenging detection of stellar binaries in the Galactic Center and in particular the identification of S-stars as stellar binaries will provide valuable information on both the formation of the S-stars and the dynamical processes in the Galactic Center.  

Here, we consider the probability for the stable binaries to eclipse, which can help distinguish them from the single star systems. We use SO-2 as an example to calculate its eclipse probability, assuming it is a binary. We find that it has a $\sim 30-40\%$ probability to eclipse at a semi-major axis of $0.1$ AU, where the general relativistic precession can suppress the eccentricity excitation of the binary due to eccentric Kozai-Lidov oscillations. The probability can be higher if the stellar binary is highly inclined relative to SO-2's orbit. We discuss the stable orbital parameters of the binary SO-2 in section \textsection \ref{s:GC}. Then, we present the calculation of the eclipse probability in section \textsection \ref{s:ep}, and the observational predictions in section \textsection \ref{s:obs}. We summarize the results and discuss the implication for the other S-stars in section \textsection \ref{s:diss}.

\section{SO-2 Binary Orbital Parameters}
\label{s:GC}
\begin{figure}[h]
\begin{center}
\includegraphics[width=3.3in]{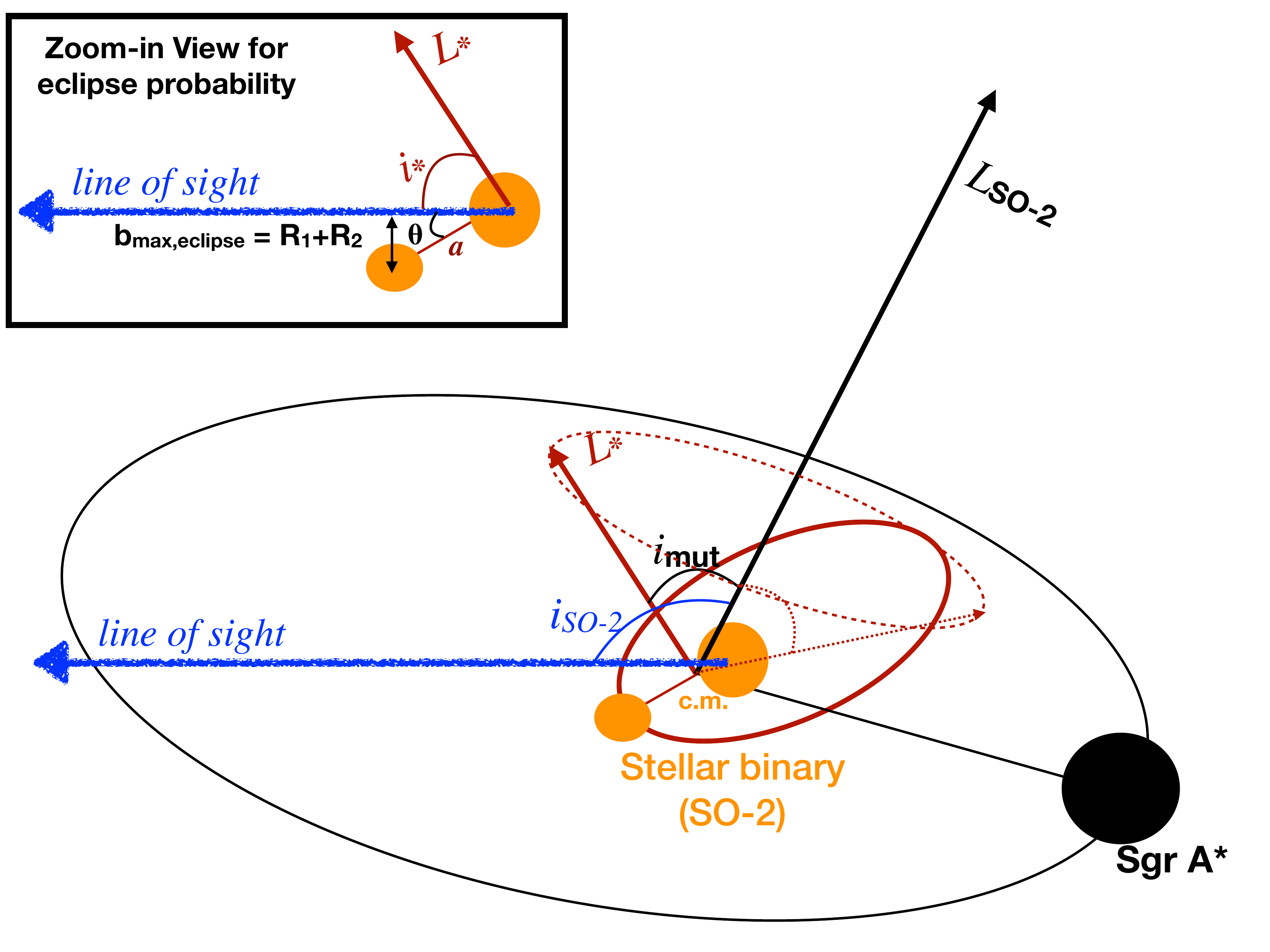} 
\caption{{\it Hypothetical binary geometry for SO-2}. `c.m.' denotes the center of mass of the binary. The inner orbit is the orbit of the binary components around the c.m., and the outer orbit is the orbit of SO-2 around Sgr A*. $i_{\rm mut}$ denotes the mutual inclination between the two orbits. The dashed cone represents the orbital spin orientation with the same mutual inclination. The blue arrow delineates the line of sight. For SO-2, the line of sight inclination is $i_{\rm SO-2} = 134.18^\circ$ \citep{Gillessen16}. \label{fig:config}}
\end{center}
\end{figure}

Binaries in the Galactic Center are difficult to resolve and distinguish from single stars. In this article, we address the probability for the binary components to eclipse each other, which may help in detection. As a proof-of-concept, we adopt SO-2 as an example, since it has the best orbital constraints. The system configuration is shown in Figure \ref{fig:config}. We use the subscript `SO-2' to denote the outer orbit of SO-2 around Sgr A*. Following \citet{Gillessen16}, we set the semi-major axis $a_{\rm SO-2} = 1000$ AU, eccentricity $e_{\rm SO-2} = 0.9$, the mass of SO-2 to be $15 {\rm M_{\odot}}$ and the mass of Sgr A* to be $4\times 10^6{\rm M_{\odot}}$. $a < 1.08$ AU is needed for the stellar binary to reside within their Hill Sphere to avoid disruption in a circular orbit with eccentricity e = 0, whereas $a<0.54$ AU when $e\to1$.

Perturbations induced by Sgr A* generally excite the eccentricity of the stellar binary and can lead to mergers of the stars \citep[e.g.,][]{Stephan16}. For stellar binaries at the location of SO-2, we find that binaries survive when the semi-major axis is $a = 0.1$ AU and $e \sim 0$, where the GR precession can suppress the Kozai-Lidov oscillations. 

\begin{figure}[h]
\begin{center}
\includegraphics[width=3.3in]{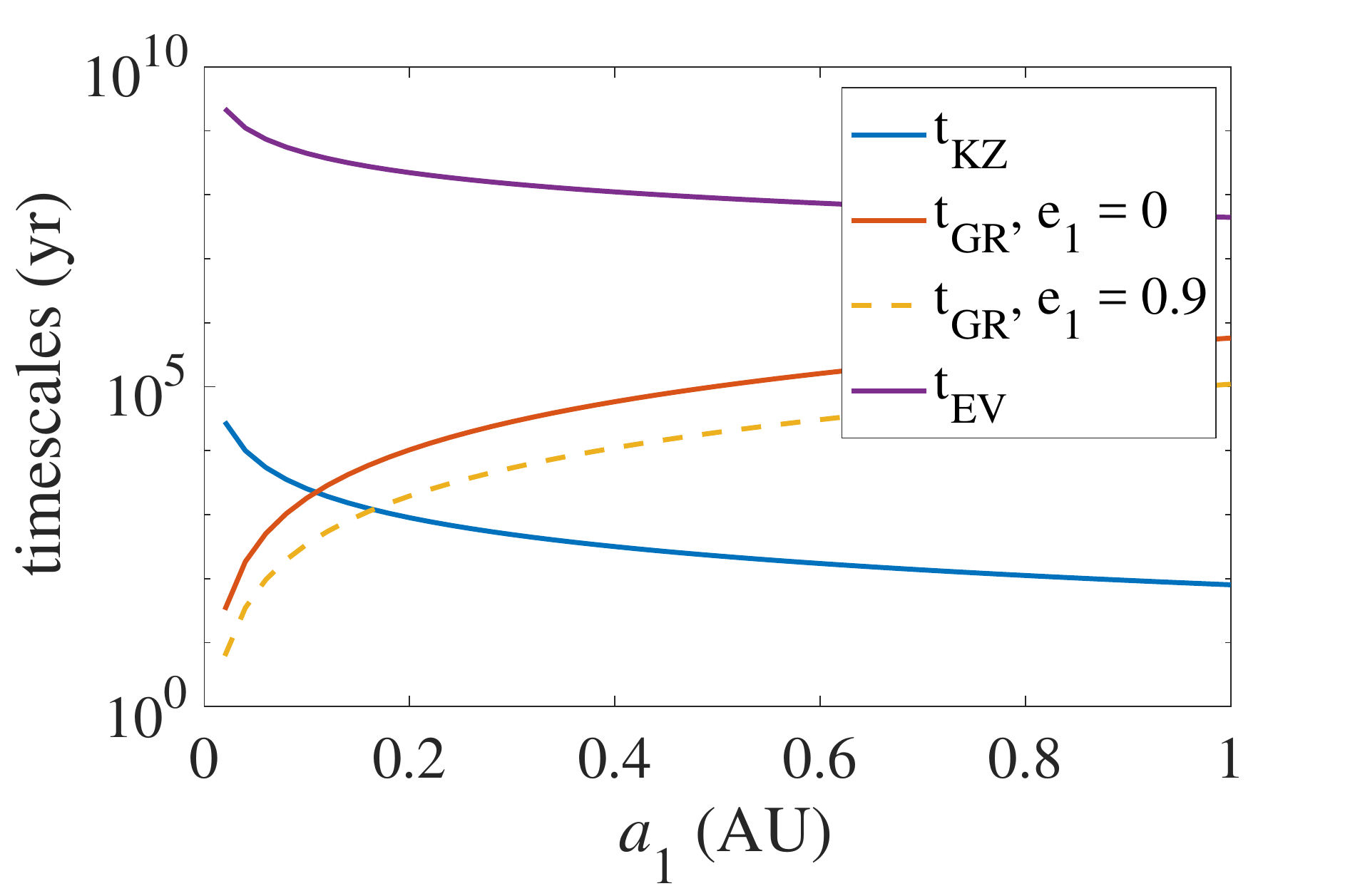} 
\caption{{\it The precession timescales ($t_{\rm GR}$ and $t_{\rm KZ}$) and the evaporation timescale ($t_{\rm EV}$) for SO-2}. The GR precession timescale ($t_{\rm GR}$) is shorter than the KZ oscillation timescale ($t_{\rm KZ}$) when $a\lesssim 0.15$ AU. The evaporation timescale is much longer than the precession timescales for small $a$. Eccentricity excitation will be suppressed when $t_{\rm KZ} < t_{\rm GR}$. \label{fig:timescale}}
\end{center}
\end{figure}

Figure \ref{fig:timescale} shows the relevant timescales of the SO-2 system. The blue line represents the Kozai-Lidov timescale; the red and the yellow dashed lines represent the precession timescales of the stellar binary due to GR with orbital eccentricities of $e = 0$ and $e=0.9$ separately (\citealt{Naoz16}). The stellar binaries could be dissociated (evaporated) through stellar encounters in the Galactic Center  \citep{Binney87}. The evaporation timescale is shown by the purple line. The Kozai-Lidov oscillations can be suppressed when the GR precession timescale is shorter, and the stellar binary is stable against dissociation within the evaporation timescale. The relevant equations can be found in \citealt{Naoz16} and \citealt{Stephan16}.

Figure \ref{fig:timescale} shows that Kozai-Lidov oscillations can be suppressed when $a\lesssim 0.15$ AU, and stars with a larger semi-major axis can survive when the orbit is more eccentric. The evaporation timescale is much longer than the precession timescales and the lifetime of the stars for small stellar binary orbital semi-major axis. As an illustrative example, we consider the case with $a = 0.1$ AU in the following discussion. 

\section{SO-2 Eclipse Probability}
\label{s:ep}

\begin{figure}[h]
\begin{center}
\includegraphics[width=3.3in]{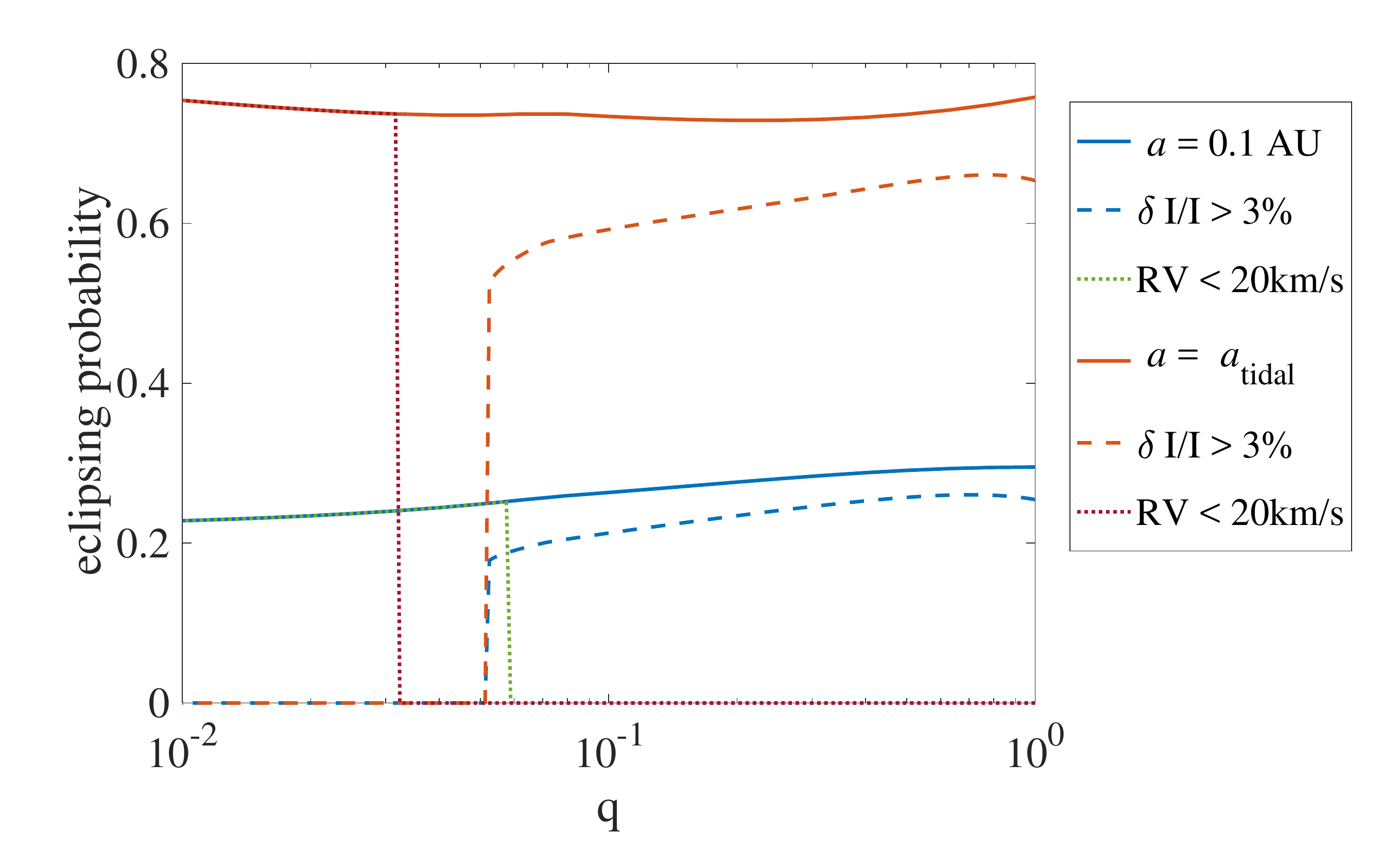} 
\caption{{\it Eclipse probability of SO-2}. The eclipse probability reaches a maximum of $\sim 75\%$ when the binary components are at their tidal disruption separation. The probability is $\sim 25\%$ when $a = 0.1$ AU. The probability depends weakly on the mass ratio of the binary ($q$). The dotted lines consider the constraints from radial velocity results, and the dashed lines include the limit from photometric amplitude for detection. Radial velocity results rule out stellar companion with mass ratio $q \gtrsim 0.05$ for $a_1 = 0.1$ AU and $q \gtrsim 0.02$ for $a_1$ at tidal limit, and the photometric amplitude requirement cannot detect stellar companion with mass ratio $q \lesssim 0.05$ (see discussions in section \ref{s:ep}). \label{fig:prob}}
\end{center}
\end{figure}

The eclipse probability of a circular binary can be expressed in a simple analytical expression. Specifically, in the case when the stellar binary orbit is circular and isotropically distributed ($\cos(i_*)$ uniformly distributed), the eclipse probability is $(R_1+R_2)/a$, where $R_1$ and $R_2$ are the stellar radii \citep[e.g.,][]{Pan12}, as illustrated in Figure \ref{fig:config}. Using this expression, we obtain the eclipse probability of SO-2 as a function of stellar mass ratio $q$, assuming a total mass for the stellar binary is $15 {\rm M_{\odot}}$ (shown in Figure \ref{fig:prob}). We follow the mass-radius relation by \citet{Kippenhahn12}, where the radius $R\propto M^{0.57}$ for a stellar mass $M>1{\rm M_{\odot}}$, and $R\propto M^{0.8}$ for $M<1{\rm M_{\odot}}$. In addition to setting $a = 0.1$ AU as a default, we also calculate the probability when the stellar binary components are separated by the maximum of their Roche Lobe radii (i.e., one of the binary components is at its tidal limit). We follow \citet{Eggleton83} to obtain the tidal limit. The tidal disruption separation is $\sim 0.03 - 0.04$ AU depending on the binary mass ratio, with the largest value obtained when the binary mass ratio is $\sim0.57$. 

Figure \ref{fig:prob} shows the eclipse probability for the SO-2 system. The blue solid line shows the case when the semi-major axis of the stellar binary is $0.1$ AU, and the red solid line corresponds to the case where the stars are at their tidal disruption separation. The eclipse probability reaches a maximum of $\sim 75\%$ when one of the stars is at its tidal limit. The eclipse probability is then $\sim 20 - 30\%$ for $a \sim 0.1$ AU. Overall, the eclipse probability depends linearly on the semi-major axis of the stellar binary, and is weakly dependent on the binary mass ratio. 

The radial velocity measurements of SO-2 provide critical constraints on the binarity of SO-2. In particular, \citet{Chu17} found that the uncertainty of radial velocity measurement of SO-2 is $\lesssim 20 {\rm km/s}$, and this rules out companion with masses $\sim 2 {\rm M_{\odot}}$ at 93.5 days binary period. We calculated the eclipse probability requiring that the radial velocity induced on the primary star is lower than $20{\rm km/s}$ when $a = 0.1$ AU and when the binary components are at the tidal limit. At the closer distances, one can rule out companions with even lower masses. The results are represented using dotted lines in Figure \ref{fig:prob}. An upper limit on the radial velocity measurements reduces the possibility of a near edge-on configuration and massive stellar companions, and thus reduces the eclipse probability. At semi-major axis of $0.1$ AU, the radial velocity constraint rules out the possibility to detect the eclipse of a stellar companion with mass ratio of $q \gtrsim 0.06$, and at the tidal limit, mass ratio of $q \gtrsim 0.02$ is ruled out.
 
On the other hand, when the stellar companion is too small or when the stellar binary is less edge on, the photometric amplitude during the eclipse is reduced. We calculate the photometric amplitude during the eclipse as following. We estimate the luminosity of the stars following the main sequence mass-luminosity relation \citep[e.g,][]{Salaris05}, then we calculate the maximum area overlap of the two stars during the eclipse ($\delta A$). The relative photometric amplitude decrease during the eclipse equates to $\delta A L_* / (\pi R_*^2) / (L_{*, 1}+L_{*, 2})$ (e.g., \citealt{Wilson71}). For simplicity, we ignored limb darkening or tidal distortion, which are higher order effects.

Requiring the photometric amplitude to exceed $\sim 3\%$ for a one sigma detection (private communication with Aurelien Hees and Tuan Do), the eclipse probability decreases. We represent the results using dashed lines in Figure \ref{fig:prob}, and find that one cannot detect the eclipse if the stellar companion mass ratio is $q\lesssim 0.05$. Thus, we consider $q = 0.05$ in our following discussion for $a = 0.1$ AU.

Both the eclipse probability and the orbital stability depend on the mutual inclination between the stellar binary and the orbit around Sgr A* (see Figure \ref{fig:config}). In particular, the orientation of the stellar binary determines the probability for the stars to eclipse along the line of sight, as the eclipse probability is larger when the stellar binary is more likely to cross the line of sight. For instance, for a circular orbit, the stellar binary orientation is allowed to cross the line of sight when $i_{\rm mut} \gtrsim i_{SO-2} - 90^\circ$, following the dashed cone in Figure \ref{fig:config}. Dynamically, the eccentricity excitation can be stronger when the stellar binary is more inclined from the orbit around Sgr A*. Next, we calculate the probability for the stellar binary to eclipse as a function of the mutual inclination between the stellar binary and their orbit around Sgr A*. 

\begin{figure}[h]
\begin{center}
\includegraphics[width=3.3in]{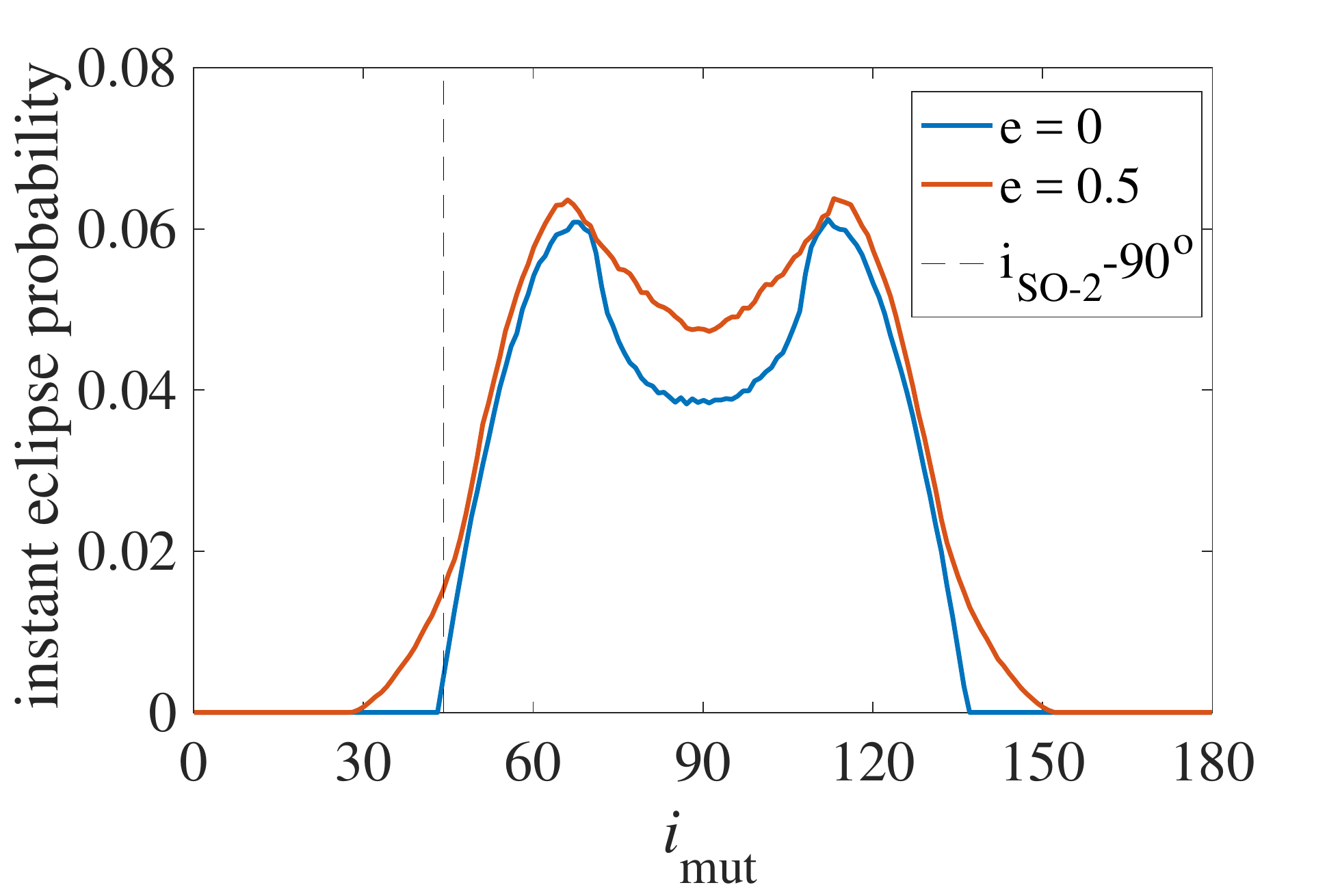} \\
\includegraphics[width=3.3in]{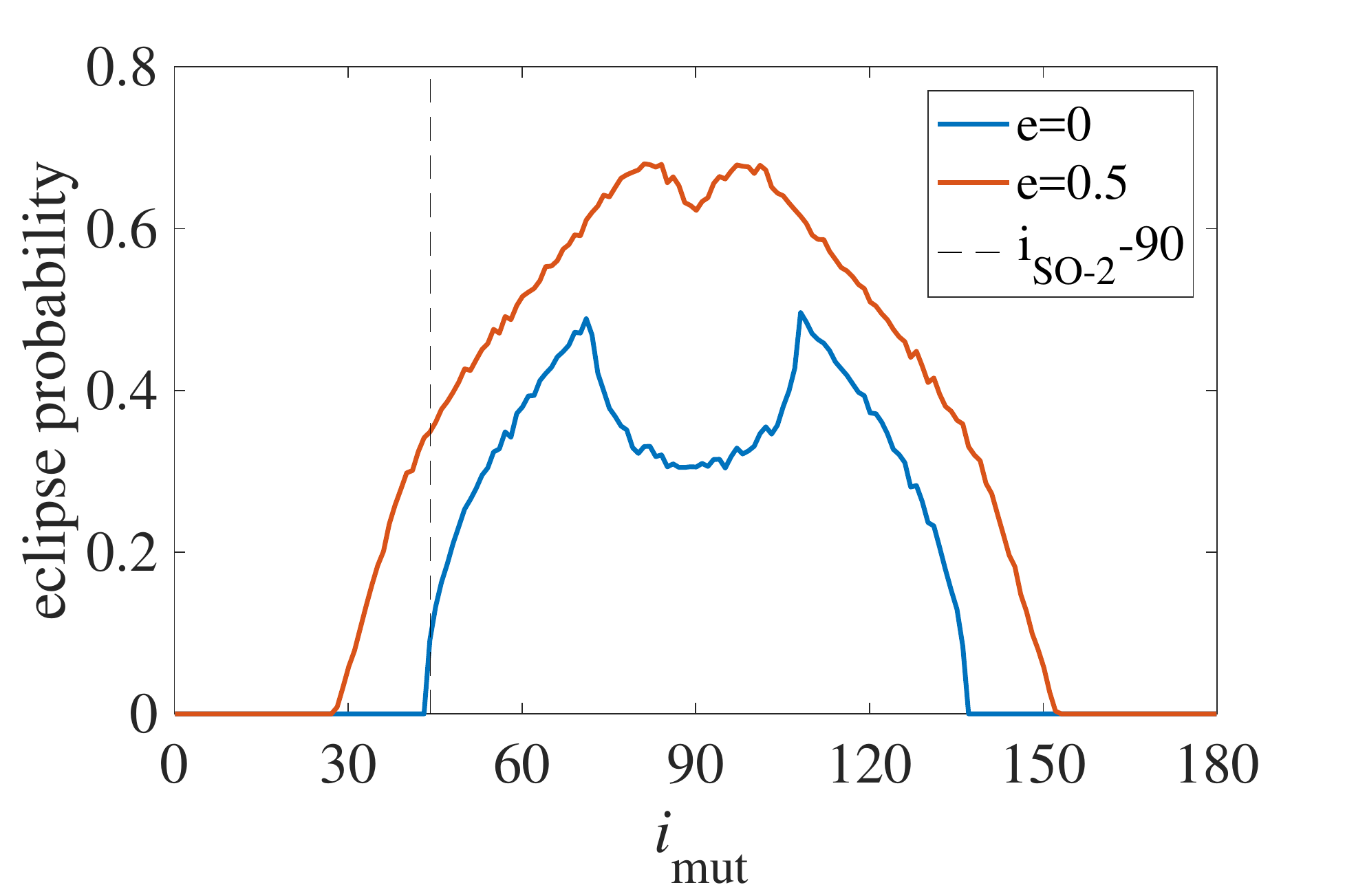} 
\caption{Upper panel: eclipse probability of SO-2 as a function of mutual inclination between SO-2 and orbit around Sgr A*. Lower panel: eclipse probability of SO-2 over the orbital period as a function of the mutual inclination. We assume $a = 0.1$ AU. \label{fig:probdi}}
\end{center}
\end{figure}

Figure \ref{fig:probdi} illustrates the eclipse probability for SO-2 as a function of the mutual inclination, where we assume the semi-major axis of the stellar binary is $0.1$ AU. \citet{Gillessen16} obtained updates on the orbit of SO-2, and found $e_{\rm SO-2} = 0.8839\pm0.0019$, $i_{\rm SO-2} = 134.18^\circ\pm0.40$, the longitude of ascending node $\Omega_{\rm SO-2} = 226.94^\circ\pm0.60$, argument of pericenter, $\omega_{\rm SO-2} = 65.51^\circ\pm0.57$, and this is consistent with the recent characterization of the orbit of SO-2 by \citet{Boehle16}, where $e_{\rm SO-2} = 0.890\pm0.005$, $i_{\rm SO-2} = 134.7^\circ\pm0.9$, $\Omega_{\rm SO-2} = 227.9^\circ\pm0.8$, $\omega_{\rm SO-2} = 66.5^\circ\pm0.9$.

To calculate the eclipse probability, we set the longitude of node for SO-2 to be $\Omega_{\rm SO-2} = 226.94^\circ$, the argument of pericenter to be $\omega_{\rm SO-2} = 65.51^\circ$ and the inclination to be $i_{\rm SO-2} = 134.18^\circ$ \citep{Gillessen16}. We use Monte Carlo simulations with uniformly distributed mean anomaly, argument of pericenter and longitude of node for the stellar binary, and we included $10^5$ realizations in the Monte Carlo simulations. We require the stellar binary to overlap in the projected plane perpendicular to the line of sight for eclipses. $i_{\rm SO-2}$ determines the probability distribution as a function of mutual inclination. The stellar binary is able to eclipse, when its orbital plane crosses the line of sight. As shown in Figure \ref{fig:config}, the binary orbit orientation lies in the dashed cone with a constant $i_{\rm mut}$. Thus, eclipses are allowed when $i_{\rm mut} \gtrsim i_{\rm SO-2}-90^\circ$. The dependence on $i_{\rm mut}$ is symmetric around $i_{\rm mut} = 90^\circ$. For general stellar binaries in the Galactic Center, when $i_{bin}$ is near zero, the probability distribution is more centrally peaked around $i_{\rm mut} = 90^\circ$, and when $i_{bin} \sim 90^\circ$, the probability distribution peaks at $i_{\rm mut} = 0^\circ$ and $180^\circ$. $\Omega_{\rm SO-2}$ and $\omega_{\rm SO-2}$ do not affect the eclipse probability.

The upper panel of Figure \ref{fig:probdi} shows the probability for the binary to eclipse at any instant of time (instantaneous eclipse probability) at a random stellar binary orbital phase, and the lower panel shows the probability over one orbital period (including all orbital phases). The blue lines correspond to a circular orbit, and the red lines correspond to an eccentric orbit. We set the mass ratio to be $q = 0.05$, but the results depends weakly on the mass ratio in the allowed region for detection, as shown in Figure \ref{fig:prob}. The probability to eclipse is higher when the mutual inclination is near $40^\circ$ and $140^\circ$. This is expected, since the orbital plane of SO-2 is highly inclined in the sky plane. Specifically, the probabilities peak near $i_{\rm mut} = i_{SO-2} - 90^\circ$ and $i_{\rm mut} = 270^\circ - i_{SO-2}$, when the line of sight inclination is able to reach $90^\circ$. The peak $i_{\rm mut}$ is not exactly at $i_{SO-2} - 90^\circ$ or $270^\circ - i_{SO-2}$, because the size of the stellar binary allows a larger parameter region for eclipse when the line of sight inclination of the stellar binary is within $i_{crit} = (R_{*,1}+R_{*, 2})/r_{12}$ from $90^\circ$, where $r_{12}$ is the separation between the binary components. For an equal mass stellar binary separated by $0.1$ AU, $i_{crit} = 17^\circ$. $i_{crit}$ determines the minimum mutual inclination, $i_{\rm mut, min} = i_{SO-2} - 90^\circ - i_{crit}$, when the eclipse probability is nonzero.

In addition, as shown in Figure \ref{fig:probdi}, the instantaneous eclipse probability depends weakly on the stellar binary orbital eccentricity, since the distance averaging over the mean anomaly depends weakly on eccentricity, ($a(1+e^2/2)$). On the other hand, the eclipse probability over one orbital period is higher when the stellar orbit is more eccentric, since the eclipse probability is higher at the pericenter of the more eccentric orbit, when the distance between the binary members is shorter. Moreover, when the orbit is more eccentric, the mutual inclination associated with the peak in the probability is closer to polar ($i_{\rm mut} \sim 90^\circ$). This is because the higher mutual inclination allows a larger phase space of orbital orientations, and the shorter distance between the stellar binary permits eclipses for a larger fraction of the possible orbital orientations. Averaging over the isotropic distribution of the stellar binary orbital orientation, we find that the total probability for the stellar binary to eclipse is $29.5\%$ for a circular orbit, and is $52.4\%$ for an eccentric orbit of $e = 0.5$ given our constraints.

 %(As shown in \citet{Ginsburginprep}, a moderate fraction of systems ($\sim 5\%$) survives with larger semi-major axis (e.g., $a = 0.5$ AU), $e = 0$ and $q = 1$, since the eccentricity excitation is weaker when the stellar binary masses are the same. ) 
%Eccentricity excitation is weaker when the stellar binary masses are the same. Thus, we also consider the eclipse probability at larger distances with $q = 0.05$. The eclipse probability scales roughly linearly with the semi-major axis, dropping with larger $a$. For example, the eclipse probability is $\sim 5\%$ when $a = 0.5$ AU for large mutual inclination $60^\circ < i_{\rm mut} < 120^\circ$. The eclipse probability peaks at around $\sim 48^\circ$ and $\sim 132^\circ$, with a value of $\sim 20\%$. The orbit is unstable when the mutual inclination is higher, corresponding to the configuration that is more likely to eclipse. Thus, the total eclipse probability is lower.

\section{Observational Implications for SO-2}
\label{s:obs}
The close separation of stable stellar binaries implies a higher probability for detecting the eclipses. However, the detection of eclipsing binaries depends sensitively on the data sampling and the eclipse duration. In this section, we consider these relevant observational parameters of the eclipses, which are useful in the search for stellar binaries. Similarly to the previous section, we adopt SO-2 as an example.

\begin{figure}[h]
\begin{center}
\includegraphics[width=3.3in, height=2.5in]{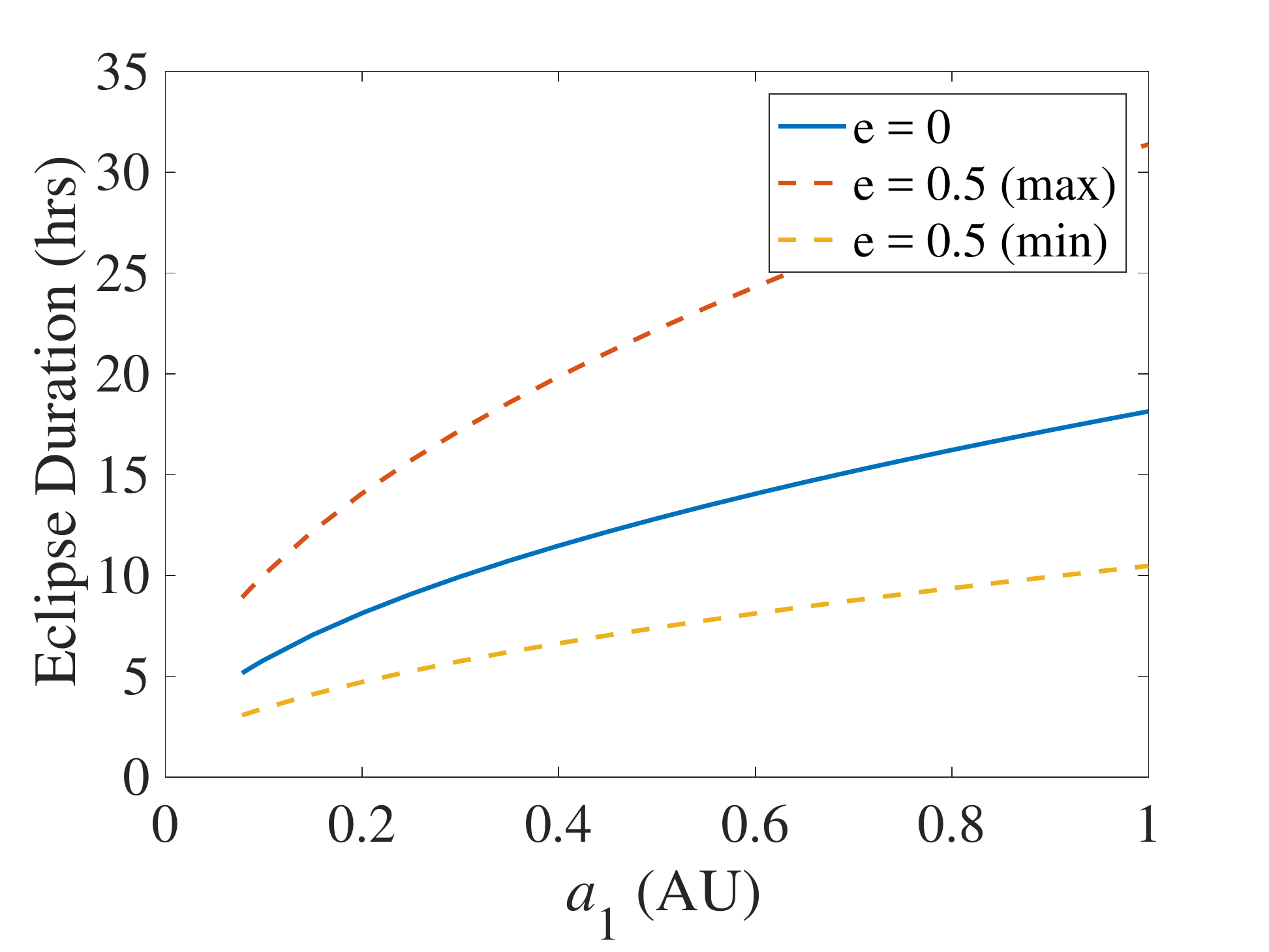} \\
\includegraphics[width=3.3in, height=2.5in]{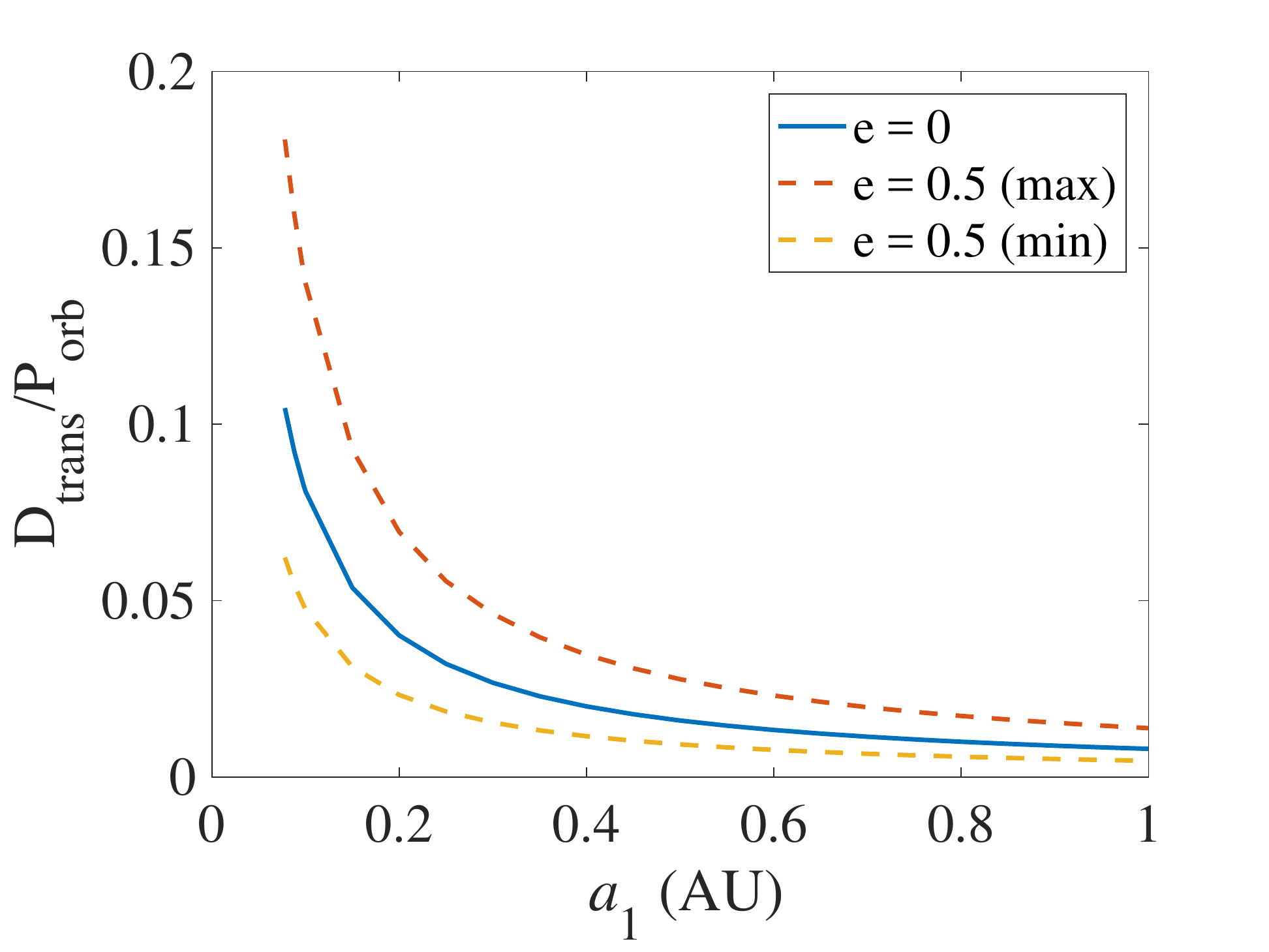} 
\caption{Upper panel: eclipse duration. Lower panel: ratio of eclipse duration to orbital period of the stellar binary. The dashed lines in the upper panel correspond to the maximum and minimum eclipse duration for an eccentric orbit ($e = 0.5$). The eclipse durations of the binaries are mostly $\lesssim$ 1 day (24 hrs) for $a\lesssim0.5$ AU. \label{fig:dtrans}}
\end{center}
\end{figure}

We first consider the duration of the eclipses, which can be written as: $P_{orb}/(\pi) {\rm asin} [\sqrt{(R_1+R_2)^2-b^2}/a]$ for a circular orbit, where $b$ is the sky projected impact parameter. We assume a zero impact parameter to estimate the duration. As shown earlier, the sum of the stellar radii depends weakly on the stellar binary mass ratio. We set $q = 0.05$ following the results shown in Figure \ref{fig:prob}. The tidal disruption separation is $0.039$ AU for the stellar binary. Thus, we consider $a>0.078$ AU for an eccentric stellar binary with $e=0.5$.

The upper panel of Figure \ref{fig:dtrans} shows the transit duration of the stellar binary as a function of semi-major axis. The transit duration is longer when $a$ increases, and the duration is lower than one day for $a \lesssim 0.5$ AU. The lower panel of Figure \ref{fig:dtrans} presents the transit duration to orbital period ratio. The eclipse duration to orbital period ratio varies from $\sim 0.1$ to $0.01$ for $a$ between $0.1 - 1$ AU.

For an eccentric orbit, we calculated the transit duration numerically. The duration varies as the orbital phase changes, with larger values when the stars are near their apo-center. The dashed lines in the upper panel of Figure \ref{fig:dtrans} illustrates the maximum and the minimum duration times for $e = 0.5$. This is obtained by uniformly sampling the binary orbit orientation and directly calculating the duration time according to the orbital phase when the stellar binaries eclipse. When the orbit is eccentric, the eclipse duration can exceed one day when $a \gtrsim 0.5$ AU. Roughly speaking, the ratio of the maximum duration time to the case when the orbit is circular is $\sim \sqrt{(1+e)/(1-e)}$, and the ratio for the minimum duration time is $\sim \sqrt{(1-e)/(1+e)}$, inversely proportional to the orbital velocity.

\section{Discussions}
\label{s:diss}
It is possible that some of the S-stars are tight stellar binaries. The close separation of the stable binary systems implies a higher eclipse probability along the line of sight. In this paper, we estimated the eclipse probability and the eclipse duration for the search of such possible binaries. Using SO-2 as an example, we found the eclipse probability is $\sim 75\%$ if the stars are at their Roche lobe limit, and $\sim 30-50\%$ if the stellar binary semi-major axis is $0.1$ AU, where the binary is stable against Kozai-Lidov oscillations. The eclipse probability is higher if the stellar binary is eccentric (reaching $\sim 50\%$ with $e = 0.5$) or when the orbit is highly misaligned relative to the orbit around Sgr A*. The eclipse duration ranges between $\sim 5 - 21$ hrs when $a$ ranges between $\sim 0.1 - 1$ AU for circular orbits, and the duration can be longer for eccentric orbits. Considering the radial velocity measurements of SO-2, stellar companion mass ratio $q \gtrsim 0.06$ can be excluded at $a=0.1$ AU. The small stellar companion produces lower photometric amplitude. Requiring the photometric amplitude to be higher than $\sim 3\%$, corresponding to a one-sigma detection limit, $q \gtrsim 0.05$ is needed. Thus, it is currently difficult to detect the eclipse of a possible low mass stellar companion of SO-2.

The S-stars in the Galactic Center provide important information on the central Supermassive Black hole (Sgr A*), and open a window to test gravitational theories \citep[e.g.,][]{Rubilar01, Hees17}. The existence of a stellar companion of the S-stars will affect the measured relativistic effects during a close approach with Sgr A* \citep{Chu17}. Thus, it is important to distinguish whether the S-stars could be binaries. In addition to SO-2, SO-175 and SO-14 also have short pericenter distances ($<50$ AU). Both the orbits of SO-175 and SO-14 around Sgr A* are nearly edge-on, i.e., $i_{SO-175} = 88.53^\circ\pm0.6$ and $100.59^\circ\pm0.87$\citep{Gillessen16}. When the stellar binary orbits are less inclined with the orbit around Sgr A* ($<40^\circ$), the stellar orbital eccentricity will less likely to be excited due to Kozai-Lidov oscillations. Then, the stable configurations of SO-175 and SO-14 allow higher probabilities to eclipse, since their binary orbits can be less inclined with the orbit around Sgr A*, and closer to the line of sight. 

\citet{Pfuhl13} searched for spectroscopic and eclipsing binaries in the Galactic Center. Selecting average photometric error $<0.1$ mag and periodic variabilities, only bright systems IRS 16SW and E60 are identified as binaries. In searching for spectroscopic binaries, only 13 brightest stars (${\rm mk} <12$) with prominent emission/absorption lines were analyzed with precise radial velocity measurements, and IRS 16NW was identified as a binary. Based on a sample of 70 stars with low photometric noise $\sigma_K \lesssim 0.04$ mag, where two of these are confirmed to be eclipsing binaries, \citet{Pfuhl13} found the fraction of eclipsing binaries is $3\pm2\%$ with photometric amplitude $>0.4$ mag. This eclipsing binary fraction depends on the eclipse duration and the photometric amplitude of the systems. Our study provides a complementary analysis on the eclipse probability assuming the S-stars are binaries, and estimate the eclipse duration of the hypothetical stellar binary of SO-2 for observational surveys. Future observations, such as with the James Webb Space Telescope (JWST) have the potential to observe near infrared sources with higher photometric precisions and help identify eclipsing binaries in the Galactic Center. The identification of stellar binaries in the S-cluster will provide valuable information on the dynamical processes leading to their formation.

\section*{Acknowledgments}
The authors are grateful to Aurelien Hees and Tuan Do for helpful discussions, and the authors would like to thank the anonymous referee, whose comments greatly improved the article. This work was supported in part by the Black Hole Initiative, which is funded by a grant from the John Templeton Foundation.




\bibliographystyle{hapj}
\bibliography{msref}

\end{document}